\documentclass[12pt,preprintnumbers,superscriptaddress,nofootinbib]{revtex4}
\usepackage{multirow}
\usepackage{amssymb}
\usepackage{amsmath, graphicx}
\usepackage{dcolumn}
\usepackage{bm}
\usepackage{enumerate}
\usepackage{slashed}
\usepackage{epstopdf}
\usepackage[unicode]{hyperref}
\usepackage{csquotes}
\usepackage[usenames]{color}

\newcommand{\be}{\begin{equation}}
\newcommand{\ee}{\end{equation}}
\newcommand{\bea}{\begin{eqnarray}}
\newcommand{\eea}{\end{eqnarray}}
\newcommand{\ben}{\begin{enumerate}}
\newcommand{\een}{\end{enumerate}}
\newcommand{\bde}{\begin{widetext}}
\newcommand{\ede}{\end{widetext}}

\newcommand{\Tr}{\mathrm{Tr}}

\newcommand{\bc}{\begin{center}}
\newcommand{\ec}{\end{center}}

\usepackage{braket}

\begin{document}

\title{\boldmath Unitary paradox of cosmological perturbations}

\author{Ngo Phuc Duc Loc}
\email{locngo148@gmail.com}

\affiliation{Department of Physics and Astronomy, University of New Mexico, Albuquerque, NM 87131, USA}

\begin{abstract}
If we interpret the Bekenstein-Hawking entropy of the Hubble horizon as thermodynamic entropy, then the entanglement entropy of the superhorizon modes of curvature perturbation entangled with the subhorizon modes will exceed the Bekenstein-Hawking bound at some point; we call this the unitary paradox of cosmological perturbations by analogy with black hole. In order to avoid a fine-tuned problem, the paradox must occur during the inflationary era at the critical time $t_c=\ln(3\sqrt{\pi}/\sqrt{2}\epsilon_HH_{inf})/2H_{inf}$ (in Planck units), where $\epsilon_H= -\dot{H}/H^2$ is the first Hubble slow-roll parameter and $H_{inf}$ is the Hubble rate during inflation. If we instead accept the fine-tuned problem, then the paradox will occur during the dark energy era at the critical time $t_c'=\ln(3\sqrt{\pi}H_{inf}/\sqrt{2}fe^{2N}H_\Lambda^2)/2H_\Lambda$, where $H_\Lambda$ is the Hubble rate dominated by dark energy, $N$ is the total number of e-folds of inflation, and $f$ is a purification factor that takes the range $0<f<3\sqrt{\pi}H_{inf}/\sqrt{2}e^{2N}H_\Lambda^2$.
\end{abstract}
\maketitle

\tableofcontents

\section{Introduction}
\begin{figure}[h!]
\centering
\includegraphics[scale=0.7]{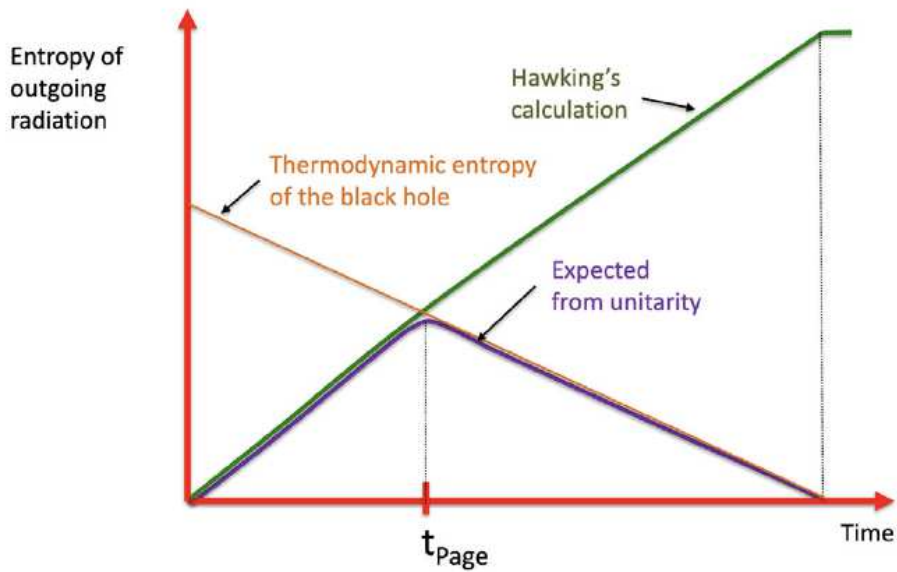}
\caption{Bekenstein-Hawking entropy versus entanglement entropy of black hole. The unitary paradox occurs at the Page time. (Figure is taken from Ref. \cite{ahmed} with permission.)}
\label{page curve}
\end{figure}

Let us first provide a brief description of the \textit{unitary paradox} of black hole. Here is the main assumption known as the \textit{central dogma}: As seen from the outside, the black hole can be described as a usual quantum system that evolves unitarily. Of course, there is a singularity at the hole's center that will require some knowledge of quantum gravity. But the singularity is hidden inside the horizon, so the black hole as seen from the outside should be reasonably considered as an ordinary quantum system. The Bekenstein-Hawking entropy of black hole is \cite{area,area2}
\begin{equation}\label{BH entropy of hole}
S_{BH}=\frac{A}{4},
\end{equation}
where $A$ is the area of the black hole's horizon \footnote{In this paper, we will use Planck units in which $\hbar=c=k_b=G=1$ (unless otherwise mentioned), so that all quantities are dimensionless.}. Bekenstein-Hawking entropy is thermodynamic entropy (also called coarse-grained entropy) of black hole and it is the maximum number of degrees of freedom (d.o.f.) available inside the hole that can be entangled with the outgoing Hawking quanta. It can also be thought of as the number of different possible configurations (or microstates) that can form the same macrostate of black hole \footnote{Strictly speaking, we should say $e^S$ d.o.f. or $e^S$ microstates (or $2^S$ microstates depending on the context). But we want to be a bit less mouthful.} . As the black hole evaporates, its horizon's area shrinks and hence $S_{BH}$ decreases asymptotically to zero \footnote{There is a caveat: When the black hole size is comparable to the Planck length, the semiclassical description is no longer reliable since we will enter the full quantum gravity regime. However, this subtlety cannot affect our arguments in any significant way as the paradox occurs well before that moment happens.}. According to Hawking's calculation, entanglement entropy\footnote{Entanglement entropy is also sometimes called Von Neumann entropy or fine-grained entropy.} of Hawking radiation grows as more and more Hawking quanta are entangled with black hole's interior d.o.f., and it saturates at a value higher than the initial Bekenstein-Hawking entropy of the hole. But this doesn't make sense, because the entanglement entropy of Hawking radiation should be equal to the entanglement entropy of black hole (if the black hole was formed from a pure state), but the entanglement entropy of black hole cannot be higher than its thermodynamic entropy (if we believe in the central dogma and the interpretation of Bekenstein-Hawking entropy). Therefore, we expect that the entanglement entropy of Hawking radiation should only increase until the Page time, when half of the black hole's d.o.f. have gone, and then decrease afterwards (see Fig. \ref{page curve}). The entanglement entropy of Hawking radiation should follow the Page curve instead of the original Hawking's calculation. This is called the \textit{unitary paradox}. There are some calculations in Ref. \cite{ahmed}  that can produce the correct Page curve.

In this paper, we study the corresponding cosmological version of the unitary paradox. In cosmology, the physical Hubble sphere with radius $H^{-1}$ plays a role similar to black hole's horizon. The \textit{central dogma} version of cosmology is: \textit{the exterior region of the Hubble horizon can be described from the inside as a usual quantum system that evolves unitarily}. As pointed out in Ref. \cite{edgar}, the central dogma conjecture for cosmological horizon has not had a firm basis yet, but we can reasonably take this as our assumption. As pointed out in subsequent sections, if we also interpret the Bekenstein-Hawking entropy of the Hubble horizon as thermodynamic entropy, then there exists a cosmological \textit{unitary paradox} as well when the entanglement entropy of curvature perturbation exceeds the Bekenstein-Hawking bound. The goal of this paper is to estimate the \textit{critical time} when the paradox occurs.

The rest of this paper is organized as follows. We discuss about the interpretation and evolution of Bekenstein-Hawking entropy of the Hubble horizon in Sec. \ref{Sec_BH}. We then talk about entanglement entropy of curvature perturbation in Sec. \ref{Sec_EE}. We point out in Sec. \ref{no fine-tuning} that, if we do not allow fine-tuning, the unitary paradox must occur during inflation. If we instead accept the fine-tuned problem, then the unitary paradox occurs during the dark energy era, as shown in Sec. \ref{yes fine-tuning}. Discussions and outlooks are in Sec. \ref{conclu}.

\section{Bekenstein-Hawking entropy}\label{Sec_BH}

By analogy with black hole (see Eq. \ref{BH entropy of hole}), the Bekenstein-Hawking  entropy  of the Hubble horizon is
\begin{equation}
S_{BH}=\frac{A}{4}=\frac{\pi}{H^2},
\end{equation}
where $H$ is the Hubble rate. The subscript ``BH" means Bekenstein-Hawking, not black hole. We interpret the Bekenstein-Hawking entropy of the Hubble horizon as thermodynamic entropy.

There are two insightful ways to think about the Bekenstein-Hawking entropy of the Hubble horizon. First, one can draw an analogy to the case of black hole. When we throw something into the black hole, it cannot escape. The black hole's horizon area will increase and hence the Bekenstein-Hawking entropy will increase. This means that the hole's horizon now encodes more information about the hole's interior (which, of course, gives some sense of nonlocality). Similarly, if we throw something outside the Hubble horizon, it will never come back due to the superluminal expansion of the exterior region. The area of the Hubble horizon will increase (since $A\propto H^{-2}$ and $H^2\propto\rho$) and hence the Bekenstein-Hawking entropy will increase (since $S_{BH}\propto A$). This means that the Hubble horizon now encodes more information about its exterior region.

Second and related to the first point, one can note that the comoving Hubble radius is $(aH)^{-1}$. During the inflationary era (or during the dark energy era), the comoving Hubble radius decreases and hence the Bekenstein-Hawking entropy for the corresponding comoving area decreases. This means that the observer inside the Hubble horizon has less and less fraction of information about the global Universe. But during the radiation or matter dominated era, the comoving Hubble radius increases and hence the Bekenstein-Hawking entropy for the corresponding comoving area increases. This means that the observer inside the Hubble horizon has more and more fraction of information about the global Universe.

In our subsequent discussions, we prefer using the physical Hubble horizon. During the inflationary era with the Hubble rate $H_{inf}$, the Bekenstein-Hawking entropy is
\begin{equation}\label{BH_inflation}
S_{BH}^{inf}=\frac{\pi}{H_{inf}^2}.
\end{equation}
During the radiation-dominated era with the Hubble rate $H_r$, the scale factor grows as $a\propto t^{1/2}$, so the Bekenstein-Hawking entropy is
\begin{equation}
S_{BH}^{rad}=\frac{\pi}{H_r^2}=4\pi t^2.
\end{equation}
During the matter-dominated era with the Hubble rate $H_m$, the scale factor grows as $a\propto t^{2/3}$, so the Bekenstein-Hawking entropy is
\begin{equation}
S_{BH}^{mat}=\frac{\pi}{H_m^2}=\frac{9\pi t^2}{4}.
\end{equation}
During the dark energy era with the Hubble rate $H_\Lambda$, the Bekenstein-Hawking entropy is
\begin{equation}\label{BH_DE}
S_{BH}^{DE}=\frac{\pi}{H_\Lambda^2}.
\end{equation}
Note that $H_{inf}$ and $H_\Lambda$ are roughly constant.

Since entanglement entropy will be calculated in momentum space, it is necessary to interpret the Bekenstein-Hawking entropy appropriately in this context:
\begin{quote}
\textit{The Bekenstein-Hawking entropy of the Hubble horizon is conjectured to be the maximum number of available  d.o.f.  of the unobserved superhorizon modes.}
\end{quote}
This is similar to the interpretation of Bekenstein-Hawking entropy of black hole when it bounds the unobserved d.o.f. inside the hole. The difference is that, while the Hilbert space is decomposed on a spatial slice for black hole, the Hilbert space for cosmological perturbations is conveniently decomposed in momentum space based on the comparison of modes' wavelength with the size of Hubble horizon (which will be discussed in the next section). In both cases, though, the deep meaning of Bekenstein-Hawking entropy is still that it bounds unobserved d.o.f.: for black hole, it bounds unobserved d.o.f. inside the hole; for cosmological horizon, it bounds unobserved d.o.f. whose wavelength is greater than the size of the horizon.

\section{Entanglement entropy}\label{Sec_EE}
During inflation, a hypothetical inflaton field is responsible for the accelerating expansion. In comoving gauge, the inflaton field is unperturbed and all scalar perturbations are carried by the \textit{curvature perturbation} $\xi(\textbf{x},t)$ of the metric\footnote{Some people might prefer the notation $\mathcal{R}(\textbf{x},t)$.}. If we just stop at the quadratic level of the action of curvature perturbation, each perturbation mode will evolve independently with a time-dependent mass due to the expansion of the Universe. But if we consider the cubic term of the action, then there exists interaction between perturbation modes. This interaction is due to the nonlinearity of general relativity and is the minimal kind of interaction that must be present. It has been shown that this interaction is the minimal mechanism responsible for decoherence during inflation \cite{nelson}. There are many terms in the cubic action, but the most dominant term responsible for decoherence is of the form $\xi_L(\partial_i\xi_S)^2$, where $\xi_L$ is the long-wavelength mode and $\xi_S$ is the short-wavelength mode. We can decompose the total Hilbert space as $\mathcal{H}_{tot}=\mathcal{H}_{k>aH}\otimes\mathcal{H}_{k<aH}$, where $\mathcal{H}_{k>aH}$ is the Hilbert space of the subsystem containing subhorizon modes and $\mathcal{H}_{k<aH}$ is the Hilbert space of the subsystem containing superhorizon modes.

Here is the basic picture of decoherence in the early Universe. We can treat the superhorizon modes as the system of interest, and the subhorizon modes as the environment (or bath modes). Decoherence happens via the interaction between the superhorizon modes and the subhorizon modes. When the modes exit the horizon during inflation, they do not immediately become classical perturbation because the coupling of interaction, which is proportional to the slow-roll parameters, is very small. It took a few e-folds of interaction for the system to be decohered and becomes classical. Because of the interaction, the superhorizon modes (the system) becomes entangled with the subhorizon modes (the environment). The entanglement entropy per physical volume of the superhorizon modes entangled with the subhorizon modes during inflation was calculated to be \footnote{As a matter of illustration, here we assumed that $\ln(\lambda^2)\sim O(1)$, where $\lambda$ is the coupling of the interacting Hamiltonian. Also note that we are using the  Planck units with $\hbar=c=k_B=G=1$, so the reduced Planck mass is $M_{pl}=m_{pl}/\sqrt{8\pi}=1/\sqrt{8\pi}$.} \cite{brahma}
\begin{equation}\label{VN_density}
s_{ent}\sim \frac{\epsilon_HH_{inf}^2a^2}{\sqrt{8\pi}},
\end{equation}
where $H_{inf}$ is roughly constant during inflation, $\epsilon_H\equiv-\dot{H}/H^2$ is the first Hubble slow-roll parameter and $a\propto \exp(Ht)$ is the scale factor. The entanglement entropy grows with time as more and more modes of perturbations exit the horizon.

If we assume that the global Universe was formed from a pure state, which is a reasonable assumption but cannot be proved easily, then the entanglement entropy of superhorizon modes must be equal to the entanglement entropy of subhorizon modes as they will combine to form a pure state. With that assumption, we can convert Eq. \ref{VN_density} to the entanglement entropy by multiplying the physical volume of the Hubble sphere
\begin{equation}\label{VN}
S_{ent}=s_{ent}\times\frac{4\pi}{3H_{inf}^3}=\frac{\sqrt{2\pi} \epsilon_H}{3H_{inf}}\exp(2H_{inf} t).
\end{equation}

\section{Unitary paradox without fine-tuning}\label{no fine-tuning}

\begin{figure}[h!]
\includegraphics[scale=1.1]{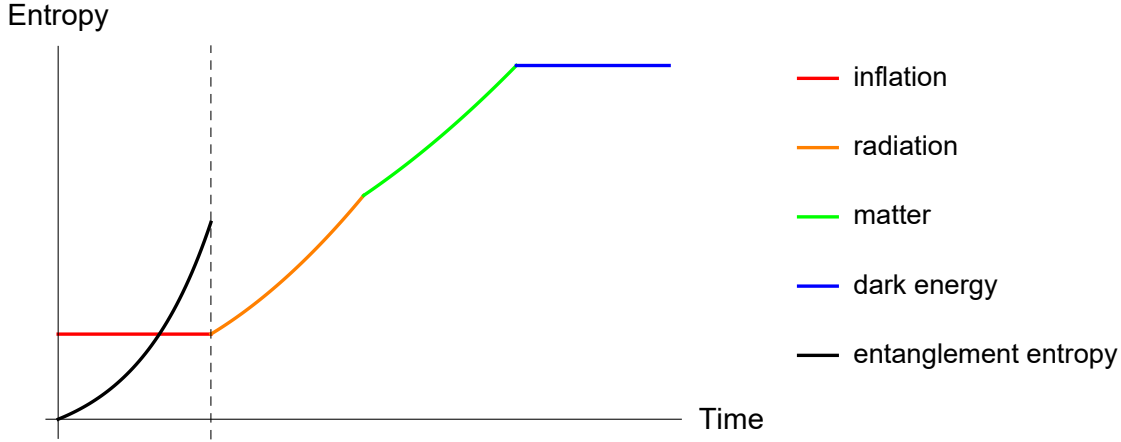}
\caption{Evolution of Bekenstein-Hawking entropy of the Hubble horizon (colorful lines) and entanglement entropy of curvature perturbations (black line). The unitary paradox occurs at the critical time $t_c$ that is given in Eq. \ref{tc}. (Figure is not to scale.)}
\label{figure}
\end{figure}

From Eqs. \ref{BH_inflation} and \ref{VN}, we can obtain the critical time at which the entanglement entropy exceeds the Bekenstein-Hawking entropy \footnote{Note that $H_{inf}$ and $\epsilon_H$ are approximately constant during inflation.}
\begin{equation}\label{tc}
t_{c}=\frac{\ln\left(3\sqrt{\pi}/\sqrt{2}\epsilon_HH_{inf} \right)}{2H_{inf}}.
\end{equation}
This is the time when the unitary paradox occurs. If we interpret the Bekenstein-Hawking entropy as the maximum number of available d.o.f. of the unobserved superhorizon modes\footnote{Note that modes must exit the horizon to be decohered and must re-enter the horizon to be observed.} , then the entanglement entropy between the subhorizon modes and superhorizon modes cannot exceed the Bekenstein-Hawking entropy because that would violate unitarity. See Fig. \ref{figure} for a sketch \footnote{Of course, in practice the Bekenstein-Hawking entropy  will grow smoothly between different eras, but as a matter of conceptual illustration we are discussing an approximate model.}.

On the one hand, the positivity condition for the above critical time\footnote{Which just means the condition for the initial entanglement entropy to be less than the Bekenstein-Hawking entropy.} is
\begin{equation}
\epsilon_H<3\sqrt{\frac{\pi}{2}}\frac{1}{H_{inf}}.
\end{equation}
If we convert this upper bound to the conventional natural units, we get $\epsilon_H<3\sqrt{\frac{\pi}{2}}\frac{m_{pl}}{H_{inf}}$. Given the fact that the typical energy scale of inflation is of order $10^{15}$ GeV and the Planck scale is of  order $10^{19}$ GeV, this upper bound is actually much milder than the usual condition $\epsilon_H<1$ for inflation to happen. In fact, the Planck's observation \cite{planck} indicates that $\epsilon_H<0.0063$ $(95\% $ C.L.), which means that our upper bound is well satisfied.

On the other hand, we note that the typical timescale of inflation is $t_{reh}\sim N/H_{inf}$, where $N$ is the total number of e-folds that is at least $\sim 70$ to solve the horizon and flatness problems \footnote{The subscript ``reh" means reheating - the moment when inflation ends.}. Therefore, the critical time happens before reheating if $t_c<t_{reh}$, which implies
\begin{equation}
\epsilon_H>3\sqrt{\frac{\pi}{2}}\frac{1}{e^{2N}H_{inf}}.
\end{equation}
 We see that this lower bound is exponentially suppressed. Although we expect that the slow-roll parameter $\epsilon_H$ should be less than unity to achieve inflation, it is unlikely that this parameter can have such a very tiny value that can violate this lower bound. Because $\epsilon_H=-\dot{H}/H^2$ changes very slowly during inflation, it must change abruptly near the end of inflation to reach $\epsilon_H=1$ to terminate inflation. In this sense, if we do not allow such a fine-tuned problem, the paradox must occur during inflation.

\section{Unitary paradox with fine-tuning}\label{yes fine-tuning}

If we instead accept the fine-tuned problem, the paradox will still occur but in the dark energy era. The real reason for this possibility is that inflation has to end to set the stage for subsequent epochs. The entanglement entropy at the end of inflation is
\begin{equation}
S_{ent}(t=t_{reh})=\frac{\sqrt{2\pi}}{3}\frac{e^{2N}}{H_{inf}}.
\end{equation}
During the radiation and matter dominated eras, the modes of cosmological perturbation re-enter the horizon and \textit{purify} with the modes already inside. We therefore expect that entanglement entropy will decrease. We parameterize this purification process with a purification factor called $f$. We absorbed all the details into this purification factor to discuss the relevant conceptual viewpoint. The entanglement entropy at the beginning of dark energy dominated era is then 
\begin{equation}
S_{ent}(t=t_{DE})=f\frac{\sqrt{2\pi}}{3}\frac{e^{2N}}{H_{inf}}.
\end{equation}
During the dark energy dominated era, the modes of perturbations exit the horizon again and entanglement entropy increases again. We can assume reasonably that it grows as $\propto \exp(2H_\Lambda t)$ similar to the case of inflation. Therefore, the entanglement entropy during the dark energy era is
\begin{equation}\label{VN_DE}
S_{ent}(t>t_{DE})=f\frac{\sqrt{2\pi}}{3}\frac{e^{2N}}{H_{inf}}\exp(2H_\Lambda t).
\end{equation}
From Eqs. \ref{BH_DE} and \ref{VN_DE}, the paradox occurs at the critical time
\begin{equation}
t_c'=\frac{\ln(3\sqrt{\pi}H_{inf}/\sqrt{2}fe^{2N}H_\Lambda^2)}{2H_\Lambda},
\end{equation}
where the purification factor $f$ satisfies
\begin{equation}
0<f<3\sqrt{\frac{\pi}{2}}\frac{H_{inf}}{e^{2N}H_\Lambda^2}.
\end{equation}
The lower bound is trivial, while the upper bound comes from the fact that $S_{ent}(t=t_{DE})<S_{BH}^{DE}$. We can see that, besides the fine-tuned problem of $\epsilon_H$ discussed earlier, we also have a fine-tuned problem for the purification factor when its upper bound is exponentially suppressed.

\section{Discussions and Outlooks}\label{conclu}
Black hole physics contains lots of paradoxes that can give us useful lessons towards a more complete and profound picture of our (effective) reality. It seems like we are trying to put everything we have on the table with the black hole at the center as a tool to connect everything we know in a consistent manner, and also to guess what are still missing. Now we are also trying to see how the whole Universe itself helps in the context of cosmology/black hole correspondence.

As seen from the outside, black hole behaves like an ordinary quantum system evolving unitarily. Because the Hubble horizon is like an inverted black hole pushing things away from each other, we can think of it as an ordinary quantum system as seen from the inside. In momentum space, the Bekenstein-Hawking entropy of the Hubble horizon is interpreted as the maximum number of available  d.o.f. of the unobserved superhorizon modes. With those observations, we pointed out that there is a corresponding unitary paradox of cosmological perturbations when entanglement entropy exceeds the Bekenstein-Hawking bound. In order to avoid the fine-tuned problems of the slow-roll parameter $\epsilon_H$ and the purification factor $f$, the paradox must occur during the inflationary era at the critical time $t_c=\ln(3\sqrt{\pi}/\sqrt{2}\epsilon_HH_{inf})/2H_{inf}$ (in Planck units). If we instead accept the fine-tuned problem, then the paradox will occur during the dark energy era at the critical time $t_c'=\ln(3\sqrt{\pi}H_{inf}/\sqrt{2}fe^{2N}H_\Lambda^2)/2H_\Lambda$, where the purification factor $f$ takes the range $0<f<3\sqrt{\pi}H_{inf}/\sqrt{2}e^{2N}H_\Lambda^2$. We remind the readers the list of major assumptions that led to these conclusions: (i) The central dogma of cosmological horizon is correct; (ii) Bekenstein-Hawking entropy of the Hubble horizon is thermodynamic entropy, which means that it is the maximum number of available d.o.f. of superhorizon modes; (iii) The global state of the Universe is pure. What is then the solution?

\begin{figure}[h!]
\includegraphics[scale=0.8]{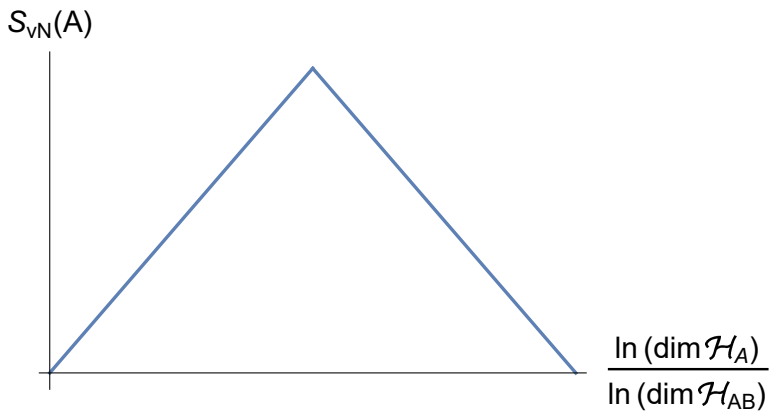}
\caption{The Page curve reaches a maximum at the point with coordinate $\{1/2,\ln(\dim \mathcal{H}_{AB})/2\}$.}
\label{page}
\end{figure}

Consider a total Hilbert space $\mathcal{H}_{AB}$ with two subsystems A and B with Hilbert space $\mathcal{H}_A$ and $\mathcal{H}_B$ respectively. The state in $\mathcal{H}_{AB}$ is pure, but the states in $\mathcal{H}_A$ and $\mathcal{H}_B$ can be entangled with each other. From the definition of entanglement entropy
\begin{equation}
S_{vN}(\rho)=-\Tr\left(\rho\  \ln\ \rho\right),
\end{equation}
it is clear that $S_{vN}(A)\leq \ln(\dim \mathcal{H}_A)$ and similarly for system B with equality if and only if all $\rho_n$ are equal, where $\rho_n$ are the eigenvalues of the density matrix $\rho$. Due to the entanglement between A and B, we also have $S_{vN}(A)=S_{vN}(B)\leq \ln(\dim\mathcal{H}_B)$. In the limit of large Hilbert space, entanglement entropy of subsystem A follows the \textit{Page curve} \cite{page,mathew}
\begin{equation}
S_{vN}(A)\sim \min\{\ln(\dim\mathcal{H}_A),\ln(\dim\mathcal{H}_B)\},
\end{equation}
which is sketched in Fig. \ref{page}. A quick qualitative understanding of this result is the following. If the size of Hilbert space of subsystem A is smaller than that of B, then $S_{vN}(A)\sim\ln(\dim\mathcal{H}_A)$  because all d.o.f. in A are maximally entangled with B, but there are still redundant d.o.f. in B that are not entangled with A. When the size of Hilbert space of A increases, more and more d.o.f. of A are entangled with B and hence $S_{vN}(A)$ increases until it reaches a maximum when $\dim\mathcal{H}_A=\dim\mathcal{H}_B$. If the size of subsystem A continues to increase, then $S_{vN}(A)$ decreases as there are more and more redundant d.o.f. in A that are not entangled with B, so we have $S_{vN}(A)\sim\ln (\dim\mathcal{H}_B)$ and the Page curve is realized. This is a general feature of entanglement between two subsystems of a random pure state with a large Hilbert space.

In the case of black hole or cosmological horizon, we also expect to have this Page-like curve but we will have to justify thoroughly for its origin since black hole and cosmological horizon are nontrivial quantum systems with classical general relativity is present. There are some directions to go from here. First, one can try to put the cosmological dogma conjecture on a firmer basis. Second, one can do detailed calculations for entanglement entropy during the radiation and matter dominated eras to get the precise value of the purification factor $f$ (if one feels okay with fine-tuning.). Third and most importantly, one can try to find a new method to compute the correct entanglement entropy that does not violate unitarity. We leave these for future works.

\textit{---Notes added:} After we submitted our paper, we were directed to Ref. \cite{10} that also discusses similar things. However, there are 3 main improvements/differences in our paper: 
\begin{enumerate}
\item The author of \cite{10} missed an important assumption that the global state of the Universe must be pure. The entanglement entropy per physical volume was calculated by tracing over the subhorizon modes. If the global Universe was formed from a pure state, the entanglement entropy of superhorizon modes will be equal to that of subhorizon modes and one can therefore multiply by the physical Hubble volume to obtain entanglement entropy within the Hubble horizon. If the global Universe was instead formed from a mixed state, the above statement is not true. We also need the global pure state assumption to apply the Page-curve argument.
\item The author of \cite{10} estimated the upper bound of the total e-folding number $N$ by demanding that the total amount of entanglement entropy generated during inflation is less than the Bekenstein-Hawking entropy by the end of inflation. This leads to a bound that is close to the Trans-Planckian Censorship Conjecture condition: $N<\ln(M_{pl}/H)$. However, this will require the energy scale of inflation to be much smaller than the conventional value $\sim 10^{15}$ GeV to ensure that this upper bound is greater than 70, because we need at least $\sim 70$ e-folds for inflation to resolve the horizon and flatness problems. In our paper, we did not have that demand and the energy scale of inflation could be large. The entanglement entropy will therefore exceed the Bekenstein-Hawking bound well before the end of inflation and we estimated the critical time when the paradox occurs.
\item We also discussed about the possibility that the paradox can happen during the dark energy dominated era due to the purification process (which is not discussed in \cite{10}). While we are currently not aware of any specific way to ease the fine-tuned problems discussed in this paper, we want to leave the door open  for possible future investigations.
\end{enumerate}

We were also noticed of Ref. \cite{hao} in which the authors considered the unitary paradox of de Sitter space in the framework of DS/dS correspondence. While their non-perturbative approach is not directly related to the curvature perturbation discussed here, they made an interesting point of deriving the Page-like curve by constructing entanglement islands.

\section*{Acknowledgment}
The author is grateful for financial support from the Department of Physics and Astronomy at UNM through the Origins of the Universe Award. The author also thanks Prof. Rouzbeh Allahverdi for correspondence.

\end{document}